\documentclass[10pt,conference]{IEEEtran}
\usepackage{amsfonts}
\usepackage{url}
\usepackage{amsmath}
\usepackage{graphics}
\usepackage{color}

\renewcommand{\QED}{\QEDopen}

\newtheorem{defn}{Definition}
\newtheorem{theorem}{Theorem}
\newtheorem{prop}{Proposition}
\newtheorem{lemma}{Lemma}

\begin{document}

\title{Strong Consistency of the Good-Turing Estimator}

\author{\authorblockN{Aaron B.\ Wagner}
\authorblockA{
Coordinated Science Laboratory \\
Univ.\ of Illinois at Urbana-Champaign \\
and School of ECE,
Cornell University \\ 
wagner@ece.cornell.edu}
\and
\authorblockN{Pramod Viswanath}
\authorblockA{
Coordinated Science Laboratory \\
and 
ECE Dept. \\
Univ.\ of Illinois at Urbana-Champaign \\
pramodv@uiuc.edu}
\and
\authorblockN{Sanjeev R.\ Kulkarni}
\authorblockA{
EE Dept. \\
Princeton University \\
kulkarni@princeton.edu}}
\maketitle
\begin{abstract}
We consider the problem of estimating the total
probability of all symbols that appear with
a given frequency in a string of i.i.d. random
variables with unknown distribution. We focus
on the regime in which the block length is large
yet no symbol appears frequently in the string.
This is accomplished by 
allowing the distribution to change with the block 
length. Under a natural convergence assumption on
the sequence of underlying distributions, we show that 
the total probabilities converge to a deterministic
limit, which we characterize. We then show that
the Good-Turing total probability estimator
is strongly consistent.
\end{abstract}

\section{Introduction}

The problem of estimating the underlying probability distribution
from an observed data sequence arises in a variety of fields
such as compression, adaptive control, and linguistics.
The most familiar technique is to use
the empirical distribution of the data, also known
as the type. This approach has a number of virtues.
It is the maximum likelihood (ML) distribution,
and if each symbol appears frequently in the string,
then the law of large numbers guarantees that the 
estimate will be close to the true distribution.

In some situations, however, not all symbols will appear
frequently in the observed data.
One example is a digital image with the pixels themselves, rather than bits,
viewed as the symbols~\cite{Orlitsky:Unknown}. Here the size of the
alphabet can meet or exceed the total number of observed
symbols, i.e., the number of pixels in the image. Another example is 
English text. Even in large corpora, many words will appear once
or twice or not at all~\cite{Efron:Thisted}. This makes 
estimating the distribution of English words using the type 
ineffective. This problem is particularly pronounced when
one attempts to estimate the distribution of bigrams, or pairs of words, 
since the number of bigrams is evidently the square of the 
number of words.

To see that the empirical distribution is lacking as an estimator 
for the probabilities of uncommon symbols, 
consider the extreme situation in which the
alphabet is infinite and we observe a length-$n$ sequence 
containing $n$ distinct symbols~\cite{Orlitsky:Science}.
The ML estimator will assign probability $1/n$
to the $n$ symbols that appear in the string and zero 
probability to the rest. But common sense suggests that
the $(n+1)$st symbol in the sequence is very likely to be one that 
has not yet appeared. It seems that the ML estimator is 
overfitting the data. Modifications to the ML estimator such
as the Laplace ``add one'' and the Krichevsky-Trofimov 
``add half''~\cite{Krichevsky:Trofimov} have
been proposed as remedies, but these only alleviate the 
problem~\cite{Orlitsky:Science}.

In collaboration with Turing, Good~\cite{Good:Turing} proposed an estimator 
for the probabilities of rare symbols that differs
considerably from the ML estimator. The Good-Turing
estimator has been shown to work well in practice~\cite{Gale:One},
and it is now used in several application areas~\cite{Orlitsky:Science}.
Early theoretical work on the estimator focused on its 
bias~\cite{Good:Turing,Robbins:Unseen,Juang:Bias}.
Recent work has been directed toward developing confidence 
intervals for the estimates using central limit
theorems~\cite{Esty:Efficiency,Mao:Poisson} or 
concentration inequalities \cite{McAllester:COLT,Drukh:Concentration}.
Orlitsky, Santhanam, and Zhang~\cite{Orlitsky:Science}
showed that the estimator has a pattern redundancy that is
small but not optimal. None of these works,
however, have shown that the estimator is strongly consistent.

We show that the Good-Turing estimator is strongly consistent 
under a natural formulation of the problem. We consider the
problem of estimating the total probability of all symbols 
that appear $k$ times in the observed string
for each nonnegative integer $k$. For $k = 0$, this is
the total probability of the unseen symbols, a quantity that
has received particular attention~\cite{Robbins:Unseen,McAllester:Missing}.
Estimating the total
probability of all symbols with the same empirical 
frequency is a natural approach when the symbols appear
infrequently so that there is insufficient data to accurately 
estimate the probabilities of the individual symbols. 
Although the total probabilities
are themselves random, we show that under our model
they converge to a deterministic limit, which we 
characterize.
Note that if the alphabet is small and the block length
is large, then the problem effectively reduces to the 
usual probability estimation problem since it is unlikely 
that multiple symbols will have the same empirical frequency.

It is known that the Good-Turing estimator performs poorly
for high-probability symbols~\cite{Orlitsky:Science}, 
but this is not a problem
since the ML estimator can be employed to estimate the probabilities
of symbols that appear frequently in the observed string.
We therefore focus on the situation in which the
symbols are unlikely, meaning that they have probability
$O(1/n)$. We allow the underlying distributions to vary 
with the block length
$n$ in order to maintain this condition, and we assume that,
properly scaled, these distributions converge. This model
is discussed
in detail in the next section, where we also describe the
Good-Turing estimator. In Section~\ref{sumsection}, we establish
the convergence of the total probabilities.
Section~\ref{GTsection} uses this convergence result to show strong
consistency of the Good-Turing estimator. Some comments regarding 
how to estimate other quantities of interest are made
in the final section.

\section{Preliminaries}

Let $(\Omega_n,\mathcal{F}_n,P_n)$ be a sequence of probability spaces.
We do not assume that $\Omega_n$ is finite or even countable.
Our observed string is a sequence of $n$ symbols 
drawn i.i.d.\ from $\Omega_n$ according to $P_n$.
Note that the alphabet and the underlying distribution are
permitted to vary with $n$. This allows us to model the
situation in which the block length is large while the
number of occurrences of some symbols is small.

\subsection{Total Probabilities}

For each nonnegative integer $k$, let $A_k^n$ denote the
set of symbols in $\Omega_n$ that appear exactly $k$ times
in the string of length $n$. We call 
\begin{equation*}
\xi_k^n := P_n(A_k^n)
\end{equation*}
the \emph{total probability} of symbols that appear $k$ times.

Of course, for $k \ge 1$, $\xi_k^n$ is simply the sum of 
the probabilities of the symbols with frequency $k$.
On the other hand, $A_0^n$ will be uncountable if $\Omega_n$ is.

We view $\xi_k^n$ as a random probability distribution
on the nonnegative integers.  Our goal is to estimate this 
distribution.

\subsection{The Good-Turing Estimator}

The Good-Turing estimator is normally viewed as an estimator
for the probabilities of the individual symbols. Let 
$\varphi_k^n = |A_k^n|$ denote the number of
symbols that appear exactly $k$ times in the
observed sequence. The basic Good-Turing estimator 
assigns probability
\begin{equation*}
\frac{(k+1)\varphi_{k+1}^n}{n \varphi_k^n}
\end{equation*}
to each symbol that appears $k \le n-1$ times~\cite{Good:Turing}. 
The case $k = n$ must be handled separately, but this case
is unimportant to us since under our model it is unlikely
that only one symbol will appear in the string.

This formula
can be naturally viewed as a total
probability estimator since the $\varphi_k^n$
in the denominator is merely dividing the total
probability equally among the $\varphi_k^n$ symbols
that appear $k$ times. Thus the Good-Turing total probability
estimator assigns probability
\begin{equation*}
\zeta_k^n := \frac{(k+1)\varphi_{k+1}^n}{n}
\end{equation*}
to the aggregate of symbols that have appeared $k$ times for
each $k$ in $\{0,\ldots,n-1\}$.
As a convention, we shall always assign
zero probability to the set of symbols that appear $n$ times
\begin{equation*}
\zeta_n^n := 0.
\end{equation*}
Like $\xi_k^n$, $\zeta_k^n$ is a random probability distribution
on the nonnegative integers.

As a total probability estimator, $\zeta_k^n$ is not ideal. 
For one thing, $\zeta_k^n$ can be positive even when $A_k^n$
is empty, in which case $\xi_k^n$ is clearly zero.
A similar problem
arises when estimating the probabilities of individual symbols,
and modifications to the basic Good-Turing estimator 
have been proposed to avoid it~\cite{Good:Turing}. But we shall 
show that even the basic form of the Good-Turing 
estimator is strongly consistent for total probability
estimation.

\subsection{Shadows}

The distributions of the total probability,
$\xi_k^n$, and the Good-Turing estimator, $\zeta_k^n$, 
are unaffected if one relabels the symbols in
$\Omega_n$. This fact makes it convenient 
in what follows to consider the probabilities
assigned by $P_n$ without reference to the labeling
of the symbols.

\begin{defn}
Let $X_n$ be a random variable on $\Omega_n$ with
distribution $P_n$. The \emph{shadow} of $P_n$ is
defined to be the distribution of the random 
variable $P_n(\{X_n\})$.
\end{defn}

As an example, if $\Omega_n = \{a,b,c\}$ and
\begin{equation*}
P_n(\{a\}) = P_n(\{b\}) = \frac{1}{2} P_n(\{c\}) = \frac{1}{4}, \\
\end{equation*}
then the shadow of $P_n$ would be uniform over
$\{1/4,1/2\}$. If $P_n$ is itself uniform, then
its shadow is deterministic. Note that the
discrete entropy of a distribution only depends
on the distribution through its shadow. We will
write $P_n(X_n)$ as a shorthand for $P_n(\{X_n\})$
in what follows.

For finite alphabets, specifying the
shadow is equivalent to specifying the unordered
components of $P_n$, viewed as a probability vector.
This is clearly seen in the above example, since the
shadow is uniformly distributed over $\{1/4,1/2\}$
if and only if the underlying distribution has two
symbols with probability $1/4$ and one with 
probability $1/2$.

If $P_n$ has a continuous component, then the shadow
will have a point mass at zero equal to the probability
of this component. The shadow reveals nothing
more about the continuous component than its
total probability, but 
we shall have no need for such information. Indeed,
the distributions of both $\xi_k^n$ and $\zeta_k^n$
depend on $P_n$ only through its shadow.

\subsection{Unlikely Symbols}

To prove strong consistency, we assume 
that the scaled profiles, $n \cdot P_n(X_n)$,
converge to a nonnegative random variable $Y$
with distribution $Q$. This implies, in particular,
that asymptotically almost every symbol has 
probability $O(1/n)$ and therefore appears
$O(1)$ times in the sequence on average. As an
example, if $P_n$ is a uniform distribution
over an alphabet of size $n$, then
the scaled shadow, $n \cdot P_n(X_n)$, equals
one a.s.\ for each $n$ 
(and hence it converges in distribution).
More complicated examples can be constructed by
quantizing a fixed density more and more finely
to generate the sequence of distributions.

\section{Total Probability Convergence}
\label{sumsection}

Before considering the performance of the Good-Turing
estimator, we study the asymptotics of the total
probabilities themselves. Under our assumption that
the scaled shadows converge, we show that the 
total probabilities converge almost surely to 
a deterministic Poisson mixture.

\begin{prop}
\label{sumprop}
The random distribution $\xi^n$ converges to
\begin{equation*}
\lambda_k := \int_0^\infty \frac{{y}^k \exp(-y)}{k!} \; dQ(y) 
    \quad k = 0,1,2,\ldots
\end{equation*}
in $L^1$ almost surely as $n \rightarrow \infty$.
\end{prop}

We prove this result by first showing that the mean
of $\xi^n$ converges  to $\lambda$ and then proving 
concentration
around the mean. To show convergence of the mean, it
is convenient to make several definitions. Let
\begin{equation*}
g_k^n(y) = {n \choose k} \left(\frac{y}{n}\right)^k
         \left(1 - \frac{y}{n}\right)^{n - k}
\end{equation*}
and
\begin{equation*}
g_k(y) = \frac{y^k \exp(-y)}{k!}.
\end{equation*}
Since
\begin{equation*}
{n \choose k} \frac{1}{n^k} \rightarrow \frac{1}{k!} \quad \text{as
    $n \rightarrow \infty$}
\end{equation*}
and
\begin{equation*}
\left(1 + \frac{y_n}{n}\right)^n 
   \rightarrow \exp(y) \quad \text{if $y_n \rightarrow y$},
\end{equation*}
it follows that for all sequences
$y_n \rightarrow y$, $g_k^n(y_n) \rightarrow g_k(y)$. Note also
that $g_k^n(y) \le 1$ if $0 \le y \le n$ by the binomial theorem.
Let
\begin{equation*}
C^n = \{\omega \in \Omega_n: P_n(\omega) > 0\}
\end{equation*}
and note that $C^n$ is countable for each $n$. 

\begin{lemma}
\label{summean}
For all nonnegative integers $k$,
\begin{equation*}
\lim_{n \rightarrow \infty} E[\xi_k^n] = \lambda_k.
\end{equation*}
\end{lemma}
\begin{proof}
We shall show that
\begin{equation}
\label{sumexpformula}
E[\xi_k^n] = E[g_k^n(nP_n(X_n))].
\end{equation}
First consider the case $k \ge 1$. Here
\begin{align*}
\xi_k^n  & = P_n(A_k^n \cap C^n) \\
        & = \sum_{\omega \in C^n} 1(\omega \in A_k^n) P_n(\omega)
\end{align*}
so by monotone convergence
\begin{align*}
E[\xi_k^n] & = \sum_{\omega \in C^n} {n \choose k} P_n(\omega)^k
                  (1 - P_n(\omega))^{n - k} P_n(\omega) \\
         & = \sum_{\omega \in C^n} g_k^n(n P_n(\omega)) P_n(\omega)  \\
          & = E[g_k^n(nP_n(X_n)) 1(X_n \in C^n)] \\
          & = E[g_k^n(nP_n(X_n))].
\end{align*}
Next consider the case $k = 0$. Here
\begin{align*}
\xi_0^n & = P_n(A_0^n) \\
    & = P_n(A_0^n \cap C^n) + P_n(A_0^n - C^n) \\
    & = \sum_{\omega \in C^n} 1(\omega \in A_0^n) P_n(\omega) + 
         P_n(\Omega_n - C^n).
\end{align*}
So again by monotone convergence,
\begin{align*}
E[\xi_0^n] & = \sum_{\omega \in C^n} (1- P_n(\omega))^n P_n(\omega) +  
                       P_n(\Omega_n - C^n) \\
     & = \sum_{\omega \in C^n} g_0^n(nP_n(\omega)) P_n(\omega) +
          P_n(\Omega_n - C^n) \\
      & = E[g_0^n(nP_n(X_n))1(X^n \in C^n)] \\  
      & \phantom{= E[g_0^n(} + 
          E[g_0^n(nP_n(X_n))1(X_n \notin C^n)] \\
      & = E[g_0^n(nP_n(X_n))].
\end{align*}
This establishes~(\ref{sumexpformula}). 
Since
$n P_n(X_n)$ converges in distribution to $Y$, we can create a 
sequence of random variables $\{Y_n\}_{n = 1}^\infty$ such that
$Y_n$ has the same distribution as $n P_n(X_n)$ and $Y_n$ converges
to $Y$ almost surely~\cite[Theorem~4.30]{Kallenberg:Foundations:2}. Then
\begin{equation*}
g_k^n(Y_n) \rightarrow g_k(Y) \quad \text{a.s.}
\end{equation*}
Since $g_k^n(Y_n) \le 1$ a.s.,
the bounded convergence theorem implies
\begin{align*}
\lim_{n \rightarrow \infty}
   E[g_k^n(Y_n)] & = E[g_k(Y)] \\
    & = \int_0^\infty g_k(y) \; dQ(y)
    = \lambda_k.
\end{align*}
\end{proof}

\begin{lemma}
\label{sumconc}
For all nonnegative integers $k$,
\begin{equation*}
\lim_{n \rightarrow \infty} |\xi^n_k - E[\xi_k^n]| = 0 \quad \text{a.s.}
\end{equation*}
\end{lemma}

\begin{proof}
Let 
\begin{equation*}
B^n = \left\{ \omega \in \Omega_n : P_n(\omega) \ge \frac{1}{n^{3/4}} 
   \right\}
\end{equation*}
and note that $|B^n| \le n^{3/4}$.  Then let
\begin{equation*}
\tilde{\xi}_k^n = P_n(A_k^n \cap B^n),
\end{equation*}
and note that
\begin{equation*}
   |\xi^n_k - E[\xi_k^n]| \le
      \left|(\xi^n_k - \tilde{\xi}^n_k) - E[\xi_k^n - \tilde{\xi}^n_k]\right| 
        + \tilde{\xi}^n_k + E[\tilde{\xi}_k^n].
\end{equation*}
Now if we change one symbol in the underlying sequence, then
$\xi^n_k - \tilde{\xi}^n_k$ can change by at most $2/n^{3/4}$. 
By the Azuma-Hoeffding-Bennett concentration 
inequality~\cite[Corollary~2.4.14]{Dembo:LD}, it follows
that for all $\epsilon > 0$
\begin{equation*}
\Pr\left(\left|(\xi^n_k - \tilde{\xi}^n_k) - E[\xi_k^n - \tilde{\xi}^n_k]
   \right| \ge \epsilon\right) 
      \le 2 \exp\left[- \frac{\epsilon^2 \sqrt{n}}{8}\right].
\end{equation*}
Since the right-hand side is summable over $n$, this implies that
\begin{equation*}
\left|(\xi^n_k - \tilde{\xi}^n_k) - 
   E[\xi_k^n - \tilde{\xi}^n_k]\right| \rightarrow 0
  \quad \text{a.s.}
\end{equation*}
Now
\begin{equation*}
\tilde{\xi}_k^n = \sum_{\omega \in B^n} P_n(\omega) 1(\omega \in A_k^n)
\end{equation*}
so
\begin{align*}
E[\tilde{\xi}_k^n] & = \sum_{\omega \in B^n} P_n(\omega) {n \choose k}
   (P_n(\omega))^k (1 - P_n(\omega))^{n - k} \\
          & \le \sum_{\omega \in B^n} {n \choose k}
   (P_n(\omega))^k (1 - P_n(\omega))^{n - k}.
\end{align*}
But 
\begin{multline*}
{n \choose k} (P_n(\omega))^k (1 - P_n(\omega))^{n - k} \\
  = \exp\left[-n\left(H\left(\frac{k}{n}\right) + 
     D\left(\frac{k}{n}\Big|\Big|P_n(\omega)\right)\right)\right],
\end{multline*}
where $H(\cdot)$ denotes the binary entropy function and 
$D(\cdot||\cdot)$
denotes binary Kullback-Leibler divergence, both with natural 
logarithms~\cite[Theorem~12.1.2]{Cover:IT}.
For all sufficiently large $n$, $k/n < 1/n^{3/4}$, which implies
that for all $\omega \in B^n$,
\begin{equation*}
D\left(\frac{k}{n}\Big|\Big|P_n(\omega)\right) \ge 
D\left(\frac{k}{n}\Big|\Big|\frac{1}{n^{3/4}}\right).
\end{equation*}
This gives
\begin{multline*}
{n \choose k} (P_n(\omega))^k (1 - P_n(\omega))^{n - k} \\
  \le {n \choose k} \left(\frac{1}{n^{3/4}}\right)^k 
    \left(1 - \frac{1}{n^{3/4}}\right)^{n - k},
\end{multline*}
so 
\begin{equation*}
E[\tilde{\xi}_k^n] 
       \le n^{3/4} {n \choose k} \left(\frac{1}{n^{3/4}}\right)^k 
          \left(1 - \frac{1}{n^{3/4}}\right)^{n - k}.
\end{equation*}
Since
\begin{equation*}
{n \choose k} \le \frac{n^k}{k!},
\end{equation*}
this implies
\begin{equation}
\label{tildeexp}
E[\tilde{\xi}_k^n] 
   \le \frac{n^{(k+3)/4}}{k!} \left(1 - \frac{1}{n^{3/4}}\right)^{n - k}.
\end{equation}
Now the right-hand side tends to zero as $n \rightarrow \infty$, so
\begin{equation*}
\lim_{n \rightarrow 0} E[\tilde{\xi}_k^n] = 0.
\end{equation*}
In fact, the right-hand side of~(\ref{tildeexp}) is summable over
$n$. By Markov's inequality,
\begin{equation*}
\Pr(\tilde{\xi}_k^n > \epsilon) \le \frac{E[\tilde{\xi}_k^n]}{\epsilon},
\end{equation*}
this implies that $\tilde{\xi}_k^n \rightarrow 0$ a.s.
The conclusion follows.
\end{proof}

\emph{Proof of Proposition~\ref{sumprop}:}
It follows from Lemmas~\ref{summean} and~\ref{sumconc} that for each $k$,
\begin{equation*}
\lim_{n \rightarrow \infty} \xi_k^n = \lambda_k \quad \text{a.s.}
\end{equation*}
That is, $\xi^n$ converges pointwise to $\lambda$ with probability
one. The strengthening to $L^1$ convergence follows from Scheff\'{e}'s
theorem~\cite[Theorem 16.12]{Billingsley:PM}, but we shall give a 
self-contained proof since it is brief.
Observe that with probability one,
\begin{align*}
0 & = \sum_{k = 0}^\infty \left[\lambda_k - \xi_k^n\right] \\
  & =   \sum_{k = 0}^\infty \left[\lambda_k - \xi_k^n\right]^+ -
     \sum_{k = 0}^\infty \left[\lambda_k - \xi_k^n\right]^-, 
\end{align*}
where $[\cdot]^+$ and $[\cdot]^-$ represent the positive and
negative parts, respectively. Thus
\begin{equation*}
\sum_{k = 0}^\infty \left| \lambda_k - \xi_k^n\right| =
   2 \sum_{k = 0}^\infty \left[ \lambda_k - \xi_k^n\right]^+ \quad \text{a.s.}
\end{equation*}
But $[\lambda_k - \xi_k^n]^+$ converges pointwise to 0 
a.s.\ and is less than or equal to $\lambda_k$. The dominated 
convergence theorem then implies that
\begin{equation*}
\lim_{n \rightarrow \infty} \sum_{k = 0}^\infty \left[\lambda_k
   - \xi_k^n\right]^+ = 0 \quad \text{a.s.}
\end{equation*}
\mbox{ } \hfill \QED

\section{Strong Consistency}
\label{GTsection}

The key to showing strong consistency is to establish
a convergence result for the Good-Turing estimator
that is analogous to Proposition~\ref{sumprop} for
the total probabilities.

\begin{prop}
\label{GTprop}
The random distribution $\zeta^n$ converges to $\lambda$ in $L^1$ almost
surely as $n \rightarrow \infty$.
\end{prop}

The desired strong consistency follows from
this result and Proposition~\ref{sumprop}.

\begin{theorem}
The Good-Turing total probability estimator is strongly consistent, i.e.,
\begin{equation*}
\lim_{n \rightarrow \infty} \sum_{k = 0}^n
  |\xi_k^n - \zeta_k^n| = 0 \quad \text{a.s.}
\end{equation*}
\begin{proof}
We have
\begin{equation*}
\sum_{k = 0}^n |\xi_k^n - \zeta_k^n| \le 
 \sum_{k = 0}^\infty |\xi_k^n - \lambda_k| + 
   \sum_{k = 0}^\infty |\lambda_k - \zeta_k^n|.
\end{equation*}
We now let $n \rightarrow \infty$ and invoke Propositions~\ref{sumprop}
and~\ref{GTprop}.
\end{proof}
\end{theorem}

The proof of Proposition~\ref{GTprop} parallels that of
Proposition~\ref{sumprop} in the previous section. In
particular, we first show that the mean of $\zeta^n$
converges to $\lambda$ and
then establish concentration around the mean.

\begin{lemma}
\label{GTmean}
For all nonnegative integers $k$,
\begin{equation*}
\lim_{n \rightarrow \infty} E[\zeta_k^n] = \lambda_k.
\end{equation*}
\end{lemma}
\begin{proof}
We shall show that
\begin{equation}
\label{GTexpformula}
E[\zeta_k^n] = E[g_k^{n-1}((n-1)P_n(X_n))].
\end{equation}
First consider the case $k \ge 1$. Here
\begin{equation*}
\zeta_k^n = \sum_{\omega \in C^n} \frac{k+1}{n} 
        1(\omega \in A_{k+1}^n).
\end{equation*}
So by monotone convergence,
\begin{align*}
E[\zeta_k^n] & = \sum_{\omega \in C^n} \frac{k+1}{n} {n \choose k + 1} 
         (P_n(\omega))^{k+1} (1 - P_n(\omega))^{n - k - 1} \\
    & =  \sum_{\omega \in C^n} {n - 1 \choose k} 
       (P_n(\omega))^{k} (1 -  P_n(\omega))^{n - k - 1}  P_n(\omega) \\
    & =  \sum_{\omega \in C^n} g_k^{n-1}((n-1)P_n(\omega)) P_n(\omega) \\
    & = E[g_k^{n-1}((n-1)P_n(X_n))1(X_n \in C^n)] \\
    & = E[g_k^{n-1}((n-1)P_n(X_n))].
\end{align*}
Next consider the case $k = 0$. Here
\begin{align*}
\zeta_0^n & = \frac{1}{n} |A_1^n| \\
          & = \frac{1}{n} |A_1^n \cap C^n| + \frac{1}{n} |A_1^n - C^n| \\
	  & = \frac{1}{n} \sum_{\omega \in C^n} 1(\omega \in A_1^n)
	       + \frac{1}{n} |A_1^n - C^n|.
\end{align*}
Again invoking monotone convergence,
\begin{align*}
E[\zeta_0^n] & = \frac{1}{n} \sum_{\omega \in C^n} {n \choose 1}
                  P_n(\omega) (1- P_n(\omega))^{n-1} \\
             & \phantom{= \frac{1}{n} \sum_{\omega \in C^n}}
		      + P_n(\Omega_n - C^n) \\
            & =  \sum_{\omega \in C^n} 
		  g_0^{n-1}((n-1)P_n(\omega)) P_n(\omega) \\
                    & \phantom{= \sum_{\omega \in C^n}}
		       + P_n(\Omega_n - C^n) \\
      & = E[g_0^{n-1}((n-1)P_n(X_n))1(X_n \in C^n)] \\
    &   \phantom{= E[g_0} + E[g_0^{n-1}((n-1)P_n(X_n))1(X_n \notin C^n)] \\
	    & = E[g_0^{n-1}((n-1)P_n(X_n))].
\end{align*}
This establishes~(\ref{GTexpformula}). Following the reasoning
in the proof of Lemma~\ref{summean}, this implies
\begin{equation*}
\lim_{n \rightarrow \infty} E[\zeta_k^n] = E[g_k(Y)] = \lambda_k
\end{equation*}
for all $k$.
\end{proof}

\begin{lemma}
\label{GTconc}
For all nonnegative integers $k$,
\begin{equation*}
\lim_{n \rightarrow \infty} |\zeta_k^n - E[\zeta_k^n]| = 0 \quad \text{a.s.}
\end{equation*}
\end{lemma}
\begin{proof}
Observe that if we alter one symbol in the underlying
i.i.d.\ sequence, then $\zeta_k^n$ will change by at
most $2(k+1)/n$. As in the proof of Lemma~\ref{sumconc}, the
Azuma-Hoeffding-Bennett concentration 
inequality~\cite[Corollary~2.4.14]{Dembo:LD} then implies 
that
\begin{equation*}
\Pr(|\zeta_k^n - E[\zeta_k^n]| > \epsilon) \le 
    2\exp\left[- \frac{\epsilon^2 n}{8 (k+1)^2}\right].
\end{equation*}
Since the right-hand side is summable over $n$, the conclusion follows.
\end{proof}

\emph{Proof of Proposition~\ref{GTprop}:}
The result follows from
Lemma~\ref{GTmean}, Lemma~\ref{GTconc}, and Scheff\'{e}'s 
theorem~\cite[Theorem 16.12]{Billingsley:PM}
as in the proof of Proposition~\ref{sumprop}.
\hfill \QED

\section{Shadow Estimation}

Proposition~\ref{sumprop} shows that the total
probabilities converge to a deterministic limit, which
is a function of the limit of the scaled shadows, $Q$. In fact, 
the total probabilities converge to  a Poisson mixture, 
with $Q$ being the mixing distribution. The functional
form of the Poisson distribution enables us to create a 
simple function of the observed string, the Good-Turing
estimator, that has the same limit as the total 
{probabilities.~In} 
particular, we can consistently estimate 
the total probabilities 
\newpage
\noindent
without having to explicitly estimate
$Q$.

In general, such a shortcut might not be 
available. It is of interest therefore to study how to
estimate $Q$ itself from the observed string. With
an estimator for $Q$, one could create a ``plug-in'' 
estimator for other quantities of interest.

\section*{Acknowledgment}

This research was supported in part by
the Army Research Office under grant DAAD19-00-1-0466
and the National Science Foundation under grants
CAREER-0237549, ITR-0325924, and CCR-0312413.

\end{document}